\documentclass[a4paper,11pt]{article}
\usepackage{pos}

\newcommand{\kms}{$\textrm{km~s$^{-1}$}$}

\newcommand{\nb}{\textsc{NBursts}}

\title{
Detailed study of galaxies with the stellar counter-rotation phenomenon
}

\author*[a]{Gasymov Damir}
\author[a,b]{Katkov Ivan}

\affiliation[a]{Sternberg Astronomical Institute, Lomonosov Moscow State University,\\
  Universitetskij pr., 13,  Moscow, 119234, Russia}

\affiliation[b]{New York University Abu Dhabi,\\
PO Box 129188, Abu Dhabi, UAE}

\emailAdd{gasymov.df18@physics.msu.ru}

\abstract{

The process of galactic disc growing is still not fully understood.
In the majority of disk galaxies the gas and stars are located in the same plane and rotate in the same direction.
However, there are kinematically peculiar galaxies hosting two counter-rotating stellar discs.
Their origin is believed to be the result of a past event of accretion of gas followed by star formation.
By studying such galaxies we can learn how much material, when, and how, have fallen onto the progenitor galaxy.
We identified a sample of 56 counter-rotating galaxies in the MaNGA IFU survey and initiated a follow-up observing campaign at the 6-m telescope (BTA) aiming to determine the stellar population properties of both stellar discs.
Our preliminary results suggest the dichotomy of the sample of counter-rotating galaxies.
We found that most massive galaxies have extended counter-rotating disks, whose contribution to luminosity is higher than in the less massive galaxies suggestive of different evolutionary paths.

}

\FullConference{%
  The Multifaceted Universe: Theory and Observations -  2022 (MUTO2022)\\
  23-27 May 2022\\
  SAO RAS, Nizhny Arkhyz, Russia\\}

\begin{document}
\maketitle

\section{Introduction}


In normal disk galaxies, gas and stars rotate in the same plane and in the same direction. However, there are galaxies with misaligned rotation of gas and stars
and even with two stellar disks rotating in opposite directions \citep{1992ApJ...394L...9R}.
The dominant evolutionary scenario for the formation of such galaxies with kinematic misalignment or counter-rotation (CR) is the accretion of external gas with different (or opposite) direction of angular momentum.
Both cosmological filaments or galaxy mergers can be a source of material forming kinematically decoupled components.
The parameters of such components and comparison with the host galaxy body allow us to learn when and how galaxies can acquire external material and form their disks.

Recent spectroscopic surveys have shown that multi-spin galaxies are more common in the galaxy population than they appeared to be 30 years ago, when they were first discovered.
For example, out of 260 early type galaxies (ETGs) from the ATLAS3D spectral panoramic survey~\citep{Davis2011MNRAS.417..882D},
30\% have been found to have a positional angle of gas rotation that differs by more than $30^\circ$ from the stellar. As shown in \citep{Bryant2019MNRAS.483..458B}, about 11\% of the 1213 galaxies in the Australian SAMI spectral survey~\citep{2018MNRAS.481.2299S} have gas-stellar misalignment. Also, 66 galaxies with this misalignment have been found, recorded, and studied by \citep{Jin2016MNRAS.463..913J} based on the MaNGA SDSS survey data. In addition to the gas-stellar misalignment, stellar counter-rotation is now frequently found in the MaNGA SDSS survey. In \citep{2022MNRAS.511.4685X, Bevacqua2022MNRAS.511..139B} a sample of galaxies with counter-rotation from the DR16 data release is reported.

Some galaxies with kinematically misaligned gas and stellar disks are the most interesting multi-spin objects. They contain two extended stellar disks rotating in opposite directions (counter-rotation, CR), located in the main plane of the galaxy. The presence of stellar counter-rotation leads to a two-peak structure in the stellar velocity dispersion map \citep{Krajnovic2011MNRAS.414.2923K}, which can be used to search for this galaxies. Stellar CR is more common in ETGs than in the late-type \citep{Corsini2014ASPC..486...51C}. An important aspect is that morphologically or photometrically such galaxies are indistinguishable from regularly rotating galaxies, and spectroscopy is needed to detect this effect. Only about a ten galaxies have been studied in detail based on deep spectroscopic observations \citep{Coccato2011MNRAS.412L.113C, Katkov2016MNRAS.461.2068K}, which were studied by the spectral decomposition methods \citep{Katkov2011BaltA..20..453K}. The application of spectral decomposition requires spectra with a high signal-to-noise ratio (S/N) and good enough spectral resolution to decompose the spectrum into two independent components with individual stellar population properties (age and metallicity) and kinematics (velocity and velocity dispersion).




In this paper, we report on the progress of the study of a sample of counter-rotating galaxies which we identified in the MaNGA survey.

\section{Sample of counter-rotating galaxies}

\subsection{MaNGA survey}


In order to increase the number of galaxies with stellar counter-rotating disks, we used the largest spectroscopic survey to date, the Mapping Nearby Galaxy at Apache Point Observatory (MaNGA) \citep{Bundy2015ApJ...798....7B}, which is part of the Sloan Digital Sky Survey (SDSS). The final Data Release~17 (DR17) in late 2021 contains  $\sim\!10\,100$ spectral cubes of unique galaxies.
The MaNGA survey covers galaxies in a large range of stellar masses and colors \citep{Wake2017AJ....154...86W}.
Each galaxy has been observed with a bundle of fibers, each 2$''$ in size.
Every bundle contains between 19 fibers (12$''$ coverage) to 127 fibers (32$''$~coverage) \citep{Drory2015AJ....149...77D}.
The survey targets were selected so that to be covered within 1.5--2.5 effective radii by these bundles, thus providing spatial information for most of the galaxy.
Circular fibers joint into bundles cannot provide a 100\% filling factor, so a dithering technique was used. The final spectral cubes have a spatial sampling of $0.5''$, while the effective spatial resolution is $2.3$--$2.5''$ \citep{Law2015AJ....150...19L}.
MaNGA provides flux-calibrated spectral cubes ready for scientific analysis and the Data Analysis Pipeline (DAP, \citep{Westfall2019AJ....158..231W} results including 2D maps of the parameters determined by fitting spectra with \textsc{ppxf} \citep{Cappellari2017MNRAS.466..798C}.
To identify the CR galaxies and other galaxies with kinematical misalignment, we visually inspected the DAP maps extensively using our web services \url{https://manga.voxastro.org},  \url{https://ifu.voxastro.org}.

\subsection{Counter-rotation sample compilation}

\begin{figure}
\centering
    \includegraphics[width=0.9\textwidth]{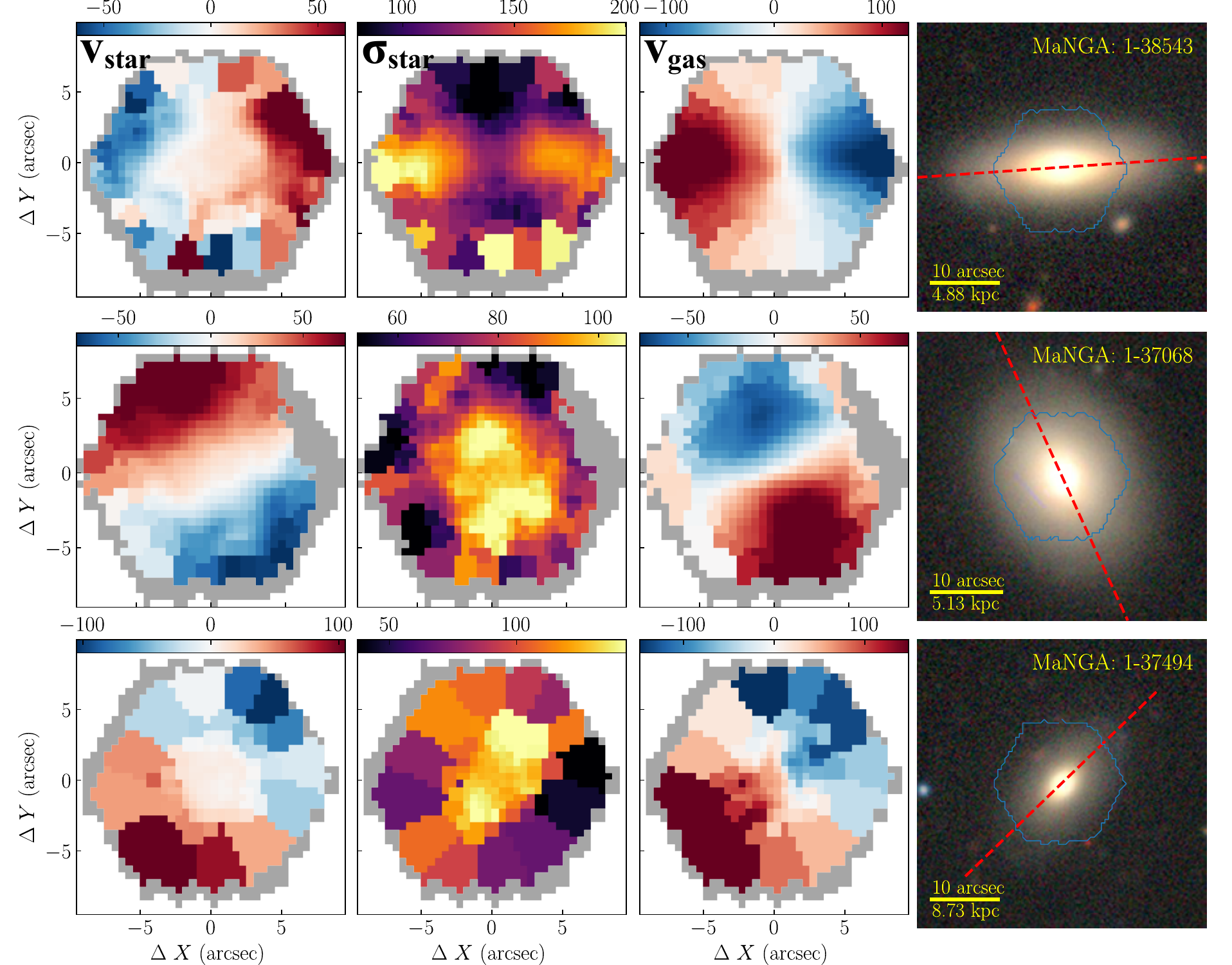}
    \caption{Kinematical maps (stellar velocity, velocity dispersion, and ionized gas velocity) for galaxies with counter-rotating stellar disks. The images were taken from the Legacy Survey. The blue and red lines show the MaNGA field of view and the SCORPIO-2 long-slit orientation respectively.}
    \label{fig:cr_image}
\end{figure}

To find galaxies with counter-rotating disks and other kinematical oddities, we checked the maps of stellar velocity, velocity dispersion, ionized gas velocities and the Legacy Survey images to make sure of disk morphology.
The main kinematic feature of stellar counter-rotation is the two off-center peaks in stellar velocity dispersion maps (\citep{Krajnovic2011MNRAS.414.2923K}, see examples in Fig.~\ref{fig:cr_image}).
In some prominent cases the central and external parts of the galaxy rotate in opposite directions, making an S-like shape of the rotation curve along the major axis. 
We also noted CR candidates with less obvious features, namely the very elongated $\sigma$-peak associated with the non-regular rotation in the central part of the galaxy, clearly demonstrating the disk morphology at the same time.


In total we identified 56 counter-rotating disk galaxies, including 30 reliable and 26 probable galaxies.
Factually, we performed the sample review in two steps, first using Data Release 16 (DR16, $\approx 4700$ targets) published in 2020 and then reviewing the new data delivered in the final DR17 ($\approx 5400$ spectra) published in late 2021.
In addition, we compiled a list of $\approx600$ galaxies with different kinematical oddities (purely gas-stellar counter-rotation, polar disks/rings, S-like rotation curve, gas-stellar misalignment).

\subsection{Analysis of sample galaxies}
\label{sec:analysis}


We analysed the MaNGA science-ready spectral cubes of the identified CR galaxies using 
the \nb\ full-spectral fitting technique \citep{nburst_a} to get kinematical and stellar ages and metallicity maps as well as best-fit models for the next step of analysis.
Note that for this part of the analysis only the galaxies identified in DR16 were used, since they were compiled earlier. We plan to perform a homogeneous analysis for the entire sample soon.

The next step in the processing is to recover the line-of-sight stellar velocity distribution (LOSVD) in a nonparametrical way.
Two counter-rotating stellar disks cause the stellar LOSVD at some position (especially along the main kinematic axis) to have a complex shape of two peaks.
Such LOSVD cannot be fitted against the Gauss-Hermite parametrization used in \nb.
Therefore, the non-parametrical LOSVD analysis is an important ingredient in the studying of the CR galaxies.
Briefly, the non-parametric LOSVD analysis can be considered as a linear problem and requires knowledge of the unbroadened galaxy spectrum, which can be taken from the first \nb\ fitting step.

The convolution of an unbroadened galaxy spectrum with an unknown LOSVD can be written as follow:
$$ Y = A \cdot \mathcal{L}~~~\longrightarrow~~~|| Y - A \cdot \mathcal{L}|| + \lambda R \cdot \mathcal{L}~\rightarrow~\text{min},$$
where matrix $A$ consists of the unbroadened spectrum shifted by one pixel, vector $Y$ is the observed galaxy spectrum at a given slit position, $\mathcal{L}$ is the desired solution. 
This problem can be solved using the standard least square method.
However, the solution is very sensitive to the noise in the data, therefore we also added regularization (matrix $R$) to the algorithm.
The regularization with a variable coefficient $\lambda$ imposes an additional condition on the solution: the smoothness of the function (L2 metric), first or second derivatives, and the tendency of the solution to zero at high speeds. A detailed description of the technique we used is presented in \citep{Gasymov2021arXiv211208386G}.

The next step is estimating a fraction of the counter-rotating stellar component based on the non-parametrically recovered LOSVD.
We analysed spectra of each individual spatial bin of the MaNGA cubes, but for simplicity we extract the 2D LOSVD along the major kinematical axis using the \textsc{pseudo-slit} package.\!\footnote{\url{https://pypi.org/project/pseudoslit/}}
The instrumental dispersion of the MaNGA spectra is as high as 69~\kms, hence the rotation (line-of-sight) amplitude must be high to identify an X-like structure (double-peaked distribution along the slit).
Of the sample of analyzed galaxies, 14 objects clearly show the X-shaped LOSVD.
Then we modelled the 2D LOSVDs 
by means of a simple,
where LOSVDs in every spatial bin was independently approximated by the sum of two Gaussians along the velocity axis, while the total flux in individual Gaussians was described by an exponential brightness profile along the slit.
As a result, it was possible to obtain a fairly accurate kinematics of the galaxy components as well as the contributions of the disks to the integral luminosity.




\section{Follow-up observations and analysis}

\begin{figure}
\centering
\includegraphics[width=\textwidth,trim=0.4cm 0.6cm 0.15cm 0.12cm,clip]{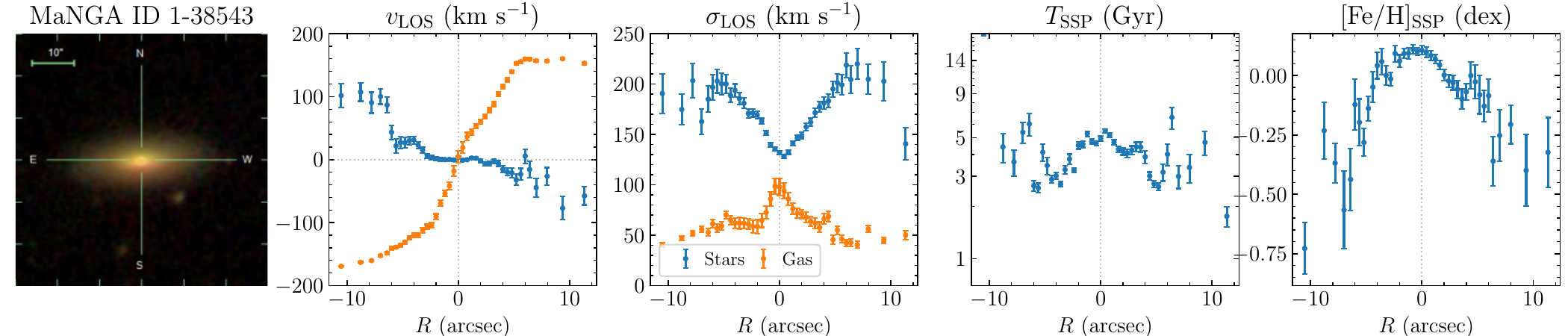}

\includegraphics[width=\textwidth,trim=0.4cm 0.6cm 0.15cm 0.12cm,clip]{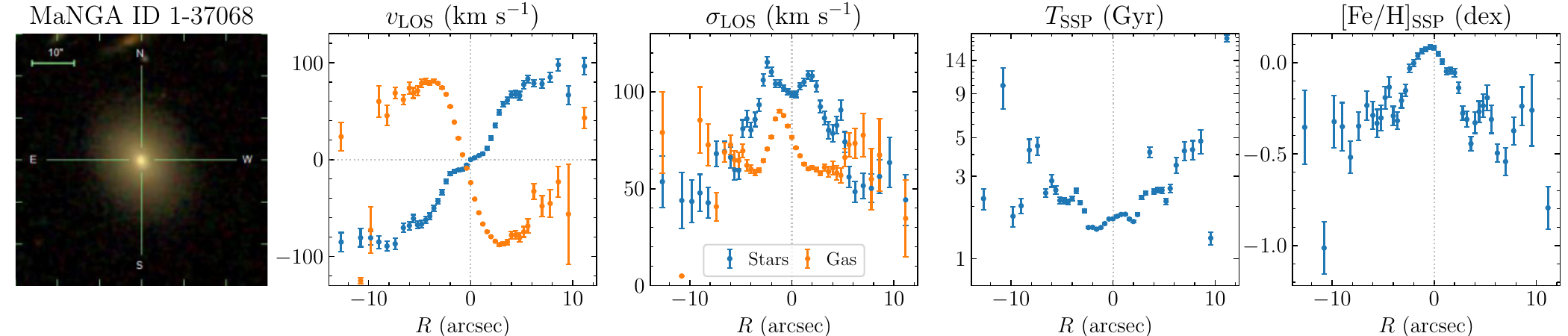}

\caption{Two examples of the radial profiles of the parameters determined from the SCORPIO-2 spectra by means of one-component \nb\ analysis.}
\label{fig:nbursts_fit}
\end{figure}

\begin{figure}
\centering
    \includegraphics[clip,trim={1cm 13cm 5.2cm 0.5cm}, height=0.154\textheight]{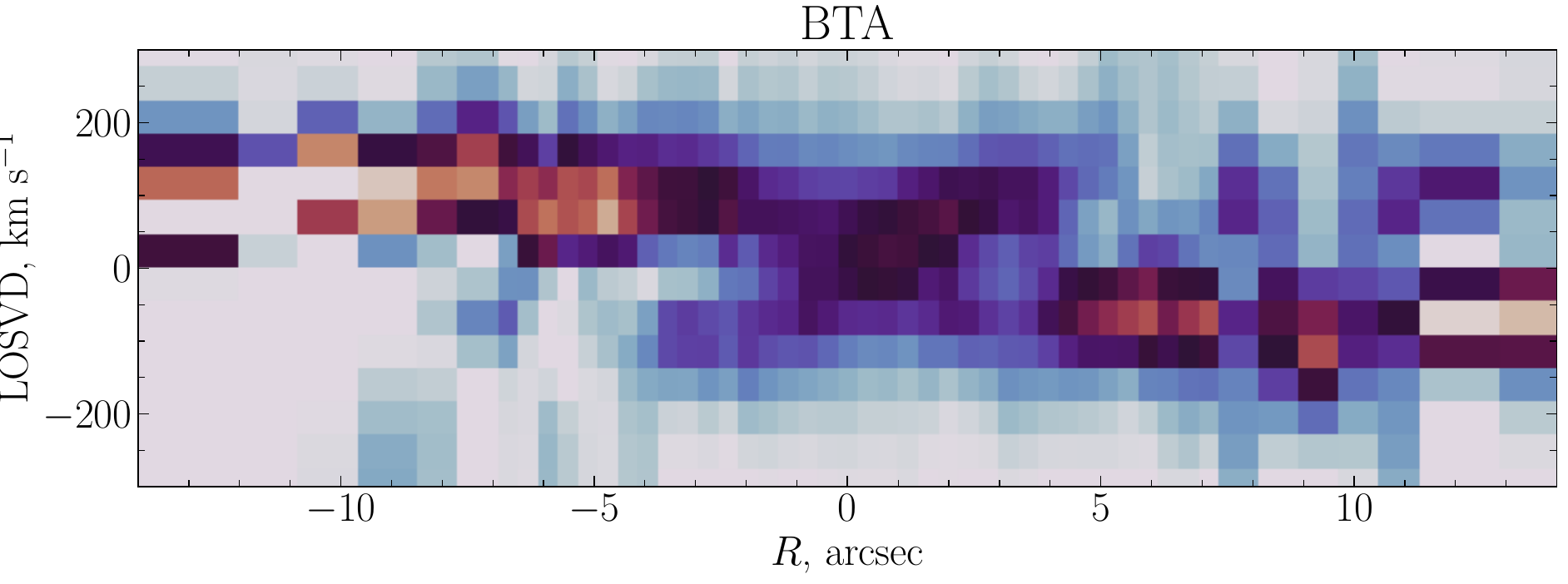}
    \includegraphics[clip,trim={9cm 0.5cm 12.6cm 13cm}, height=0.154\textheight]{figs/comp_1-115097_BTA_MaNGA.pdf}
    \caption{A comparison of the non-parametric LOSVDs obtained from the MaNGA IFU spectrum and the long-slit BTA spectrum in equal scales.}
    \label{fig:losvd_comparison_MaNGA_BTA}
\end{figure}

Only some galaxies show a clear kinematical bimodality of disks, based on the MaNGA data.
It is likely due to the relatively low spectral resolution and signal-to-noise ratio in the data.
Therefore, we initiated a follow-up spectroscopic program at the 6-m Russian telescope (BTA) to obtain deep long-slit spectra which would allow us to precisely determine the kinematical structure and stellar population properties of the counter-rotating disks.
At the moment, 8 CR galaxies have been observed using the universal spectrograph SCORPIO-2 \citep{Afanasiev2011BaltA..20..363A} (see Table \ref{tab:tab1}) with a 0.5\arcsec\ slit and a 1200@540 grism.
This mode provides an instrumental dispersion about 50~\kms, which is better than the MaNGA one  ($\approx75$~\kms).
Long total exposures (typically 2h) allowed us to obtain deep spectra, which we analyzed in the same manner as described in Section~\ref{sec:analysis}.
The examples of the radial profiles of the determined kinematical and stellar population parameters for two CR galaxies are shown in Fig.~\ref{fig:nbursts_fit}.
Thanks to high quality SCORPIO-2 spectra, the non-paramertrical LOSVD recovery and its analysis turn out better than the MaNGA results.
Fig.~\ref{fig:losvd_comparison_MaNGA_BTA} demonstrates the difference in the recovered LOSVD of MaNGA 1-115097 for the MaNGA and SCORPIO-2 spectra.
Indeed, we were able to estimate the fraction of counter-rotating stars for all the 8 galaxies, while the analysis of the MaNGA data failed for 4 targets where the CR fraction is low (see Table~\ref{tab:tab1}).


\begin{table}
\begin{center}
\scalebox{0.85}{
\begin{tabular}{|c|c|c|c|c|c|c|c|c|}
\hline
MaNGA ID & \multicolumn{2}{c|}{Age, Gyr} & \multicolumn{2}{c|}{[Fe/H], dex} & CR loc. & Mass, & W$_\textrm{BTA}$ & W$_{\textrm{MaNGA}}$ \\ 
\cline{2-5}
 & CR & Main & CR & Main &  & $10^9$ M$_\odot$ &  &  \\ 
\hline
\hline
1-115097 & 3 & 5 & 0.0 & -0.5 & inner & 26.4 & 15\% & --- \\
\hline
1-136248 & 4 & 7 & -0.3 & 0.1 & outer & 34.8 & 78\% & 81\% \\
\hline
1-248869 & 4 & 9 & -0.4 & 0.0 & outer & 105 & 49\% & 47\% \\
\hline
1-37062 & 2 & 3 & 0.1 & -0.5 & inner & 15 & $\sim$ 5\% & --- \\
\hline
1-37068 & 1.5 & 2.5 & 0.1 & -0.3 & inner & 23.7 & $\sim$ 6\% & --- \\
\hline
1-38543 & 2.5 & 5 & -0.2 & 0.1 & outer & 31.6 & 31\% & 21\% \\
\hline
1-635506 & 2.5 & 3 & 0.0 & -0.3 & inner & 16.6 & $\sim$ 5\% & --- \\
\hline
1-94690 & 2.5 & 6 & 0.0 & -0.6 & inner & 24.4 & 21\% & 14\% \\
\hline
\end{tabular}}
\small
\end{center}
    \caption{The stellar population parameters of the counter-rotating (left numbers) and old (right) disks of the galaxies observed with the BTA: the location of the counter-rotating disk, galaxy mass \citep{2018ApJ...859...11S}, contribution into  integral luminosity, determined in the non-parametric LOSVD analysis from the BTA and MaNGA spectra.}
    \label{tab:tab1}
\end{table}


\section{Summary}

In this proceeding we present the progress on our ongoing project to study galaxies with counter-rotating stellar discs identified in the MaNGA spectroscopic survey.
Visually inspecting the entire MaNGA sample, we found 56 counter-rotating galaxies, including 30 reliable and 26~probable galaxies and around 600 galaxies showing other signs of kinematical misalignment (gas-stellar counter-rotation, polar disks/rings, etc.).
We followed 8 CR galaxies up from the parent sample with the SCORPIO-2 spectrograph mounted at the 6-m telescope (BTA). 
The higher spectral resolution and signal-to-noise of the SCORPIO-2 data allowed us to estimate more accurately the light fraction of the counter-rotating stellar disks, which turned out to be in a wide range (5--80\%).
It is assumed that the later formed stellar CR component rotates in the opposite direction relative to the main body of the galaxy. First of all, our sample contains galaxies with both extended and compact CR disks, suggestive a dichotomy of properties.
We noticed that extended CR disks possess highest light fractions and inhabit the most massive galaxies.
This gives clear evidence for the dichotomy of the population of counter-rotating galaxies.

We plan to apply the spectral decomposition technique to the obtained SCORPIO-2 data in order to accurately estimate the stellar population parameters and discuss different formation scenarios, having in hand the whole set of parameters such as the light and mass fraction of CR disks, their spatial distribution, age, and metallicity.

\section*{Acknowledgements}

This project is supported by the Russian Science Foundation grant 21-72-00036.
Part of the observed data was obtained on the unique scientific facility ``The Big Telescope Alt-Azimuthal'' of SAO~RAS, and the data processing was supported under grant  075-15-2022-262 (13.MNPMU.21.0003) of the Ministry of Science and Higher Education of the Russian Federation.
We also thank Mr.~Evgenii Malygin, Mr.~Sergey Kotov and Dr.~Alexei Moiseev for making telescope observations from the paper.

\bibliographystyle{JHEP}
\bibliography{bib}
\end{document}